\documentclass[10pt,conference]{IEEEtran}
\IEEEoverridecommandlockouts
\usepackage{cite}
\usepackage{amsmath,amssymb,amsfonts}
\usepackage{algorithmic}
\usepackage{graphicx}
\usepackage{textcomp}
\usepackage{xcolor}

\newcommand\gs[1]{\mathfrak{S}_{#1}}

\newcommand\pgs[1]{\mathfrak{P}_{#1}}

\def\merge{\amalg}

\newcommand\pad{\operatornamewithlimits{\rightharpoonup}}

\newcommand\concat{\operatornamewithlimits{\mathbin{{+}\mspace{-8mu}{+}}}}

\usepackage{tikz}
\usetikzlibrary{calc,arrows,decorations.pathmorphing,intersections,patterns}
\usepackage{pgf,pgfplots} 
\usepgfplotslibrary{colormaps}

\pgfdeclarepatternformonly{north westish lines}{\pgfqpoint{-0pt}{-0pt}}{\pgfqpoint{6pt}{6pt}}{\pgfqpoint{6pt}{6pt}}{
        \pgfsetlinewidth{0.6pt}
        \pgfpathmoveto{\pgfqpoint{0pt}{0pt}}
        \pgfpathlineto{\pgfqpoint{6pt}{6pt}}
        \pgfpathmoveto{\pgfqpoint{5.8pt}{-.2pt}}
        \pgfpathlineto{\pgfqpoint{6.2pt}{.2pt}}
        \pgfpathmoveto{\pgfqpoint{-.2pt}{5.8pt}}
        \pgfpathlineto{\pgfqpoint{.2pt}{6.2pt}}
        \pgfusepath{stroke}}

\pgfdeclarepatternformonly{north eastish lines}{\pgfqpoint{-0pt}{-0pt}}{\pgfqpoint{6pt}{6pt}}{\pgfqpoint{6pt}{6pt}}{
        \pgfsetlinewidth{0.6pt}
        \pgfpathmoveto{\pgfqpoint{0pt}{6pt}}
        \pgfpathlineto{\pgfqpoint{6pt}{0pt}}
        \pgfpathmoveto{\pgfqpoint{3pt}{9pt}}
        \pgfpathlineto{\pgfqpoint{9pt}{3pt}}
        \pgfpathmoveto{\pgfqpoint{-3pt}{3pt}}
        \pgfpathlineto{\pgfqpoint{3pt}{-3pt}}
        \pgfusepath{stroke}}

\def\hatchspacing{3.775pt}
\pgfdeclarepatternformonly{hatchish lines}{\pgfqpoint{-0pt}{-0pt}}{\pgfqpoint{\hatchspacing}{\hatchspacing}}{\pgfqpoint{\hatchspacing}{\hatchspacing}}{
        \pgfsetlinewidth{0.8pt}
        \pgfpathmoveto{\pgfqpoint{0pt}{\hatchspacing}}
        \pgfpathlineto{\pgfqpoint{\hatchspacing}{\hatchspacing}}
        \pgfpathmoveto{\pgfqpoint{\hatchspacing}{0pt}}
        \pgfpathlineto{\pgfqpoint{\hatchspacing}{\hatchspacing}}
        \pgfusepath{stroke}}

\def\BibTeX{{\rm B\kern-.05em{\sc i\kern-.025em b}\kern-.08em
    T\kern-.1667em\lower.7ex\hbox{E}\kern-.125emX}}
\begin{document}

\title{Performant coherent control: bridging the gap between high- and low-level operations on hardware \\
\thanks{US Department of Energy, Office of Science, Office of Advanced Scientific Computing Research Quantum Testbed Program and National Quantum Information Science Research Centers, Quantum Systems Accelerator} 
}

\author{\IEEEauthorblockN{Daniel S. Lobser}
  \IEEEauthorblockA{\textit{Sandia National Laboratories}\\
  Albuquerque, NM, USA\\
  dlobser@sandia.gov} 
  \and
  \IEEEauthorblockN{Jay W. Van Der Wall}
  \IEEEauthorblockA{\textit{Sandia National Laboratories}\\
  Albuquerque, NM, USA \\
  }
  \and
  \IEEEauthorblockN{Joshua D. Goldberg}
  \IEEEauthorblockA{\textit{Sandia National Laboratories}\\
  Albuquerque, NM, USA \\
  }
}

\maketitle
\begin{abstract} 

Scalable coherent control hardware for quantum information platforms is rapidly growing in priority as their number of available qubits continues to increase. As these systems scale, more calibration steps are needed, leading to challenges with system instability as calibrated parameters drift. Moreover, the sheer amount of data required to run circuits with large depth tends to balloon, especially when implementing state-of-the-art dynamical-decoupling gates which require advanced modulation techniques. We present a control system that addresses these challenges for trapped-ion systems, through a combination of novel features that eliminate the need for manual bookkeeping, reduction in data transfer bandwidth requirements via gate compression schemes, and other automated error handling techniques. Moreover, we describe an embedded pulse compiler that applies staged optimization, including compressed intermediate representations of parsed output products, performs \emph{in-situ} mutation of compressed gate data to support high-level algorithmic feedback to account for drift, and can be run entirely on chip.

\end{abstract}


\section{Introduction}

As quantum computing (QC) platforms continue to scale, increasing demands are placed on hardware control capabilities. These include basic features, such as synchronous execution of pulse sequences used for realizing coherent gate operations, and extensibility so that hardware outputs can be scaled to accommodate more qubits. However, other requirements can arise as systems scale, especially for noisy intermediate-scale quantum (NISQ) devices, where increasing numbers of calibrations and drift in control parameters impact system stability and, in turn, the time one can dedicate to running meaningful quantum circuits.

These challenges can be offset by moving more classical computing tasks directly onto the control hardware. This enables techniques such as automated calibration and high-level feedback in which external computing resources are removed from the loop, cutting down on communication overhead for over-the-wire data transfers. 
Effective implementation of classical operations on chip further supports a fully-distributed architecture, where classical resources scale proportionately with hardware outputs. If compilation on the embedded system can outpace an external server (when accounting for network transfer times), this paves the way for self-contained control hardware.

In this paper, we present progress towards such a distributed architecture, by using hardware/software codesign to demonstrate compilation of quantum assembly down to the pulse level, with high-level support for pulse engineering techniques, directly on coherent control hardware. 
We further demonstrate that the embedded compiler outperforms an external server, when accounting for upload times, where we primarily target an application in which high-level shot-to-shot feedback is used to account for drift in gates that utilize pulse shaping.
\section{Background}
Control systems for quantum platforms often come with a complex set of requirements. General-purpose commercial off-the-shelf (COTS) hardware typically fails to meet platform-specific needs, requiring multiple COTS components to be cobbled together. This can introduce unnecessary challenges for dealing with imperfections in external components. To wit, external rf components introduce nonlinear coupling in calibrated parameters due to amplitude- and frequency-dependent phase shifts. Shimming out such hardware imperfections adds undue burden when working with already complicated quantum systems.

An alternative approach is to develop a streamlined control system to address the ever-evolving requirements of specific quantum platforms head on. While these requirements can be quite nuanced, and vary from system to system, customized handling of system constraints at the hardware level offers significant performance gains. Moreover, hardware-level tailoring can achieve a concise interface not possible with generalized systems by decoupling platform-dependent error mitigation techniques from a basic set of idealized operations.

We employ the latter approach, where the design is implemented on a system specially designed for the Quantum Scientific Computing Open User Testbed (QSCOUT)~\cite{b1}. While many of the details can be found in Ref.~\cite{b1}, relevant aspects of the QSCOUT apparatus, as they pertain to control hardware requirements, are described here. 

\subsection{Target system}

The experiment uses a linear array of trapped $^{171}\text{Yb}^+$ ion qubits confined to a microfabricated surface ion trap. Qubit states are controlled using optical Raman transitions via individual addressing (IA) beams. The IA beams are generated from a single pulsed laser, where the frequency comb enables Raman frequencies to be tuned directly to the $^{171}\text{Yb}^+$ 12.642 GHz clock transition using a multichannel acousto-optic modulator (AOM)~\cite{Hayes2010}. The pulsed laser is centered around 355 nm, which is conveniently near a minimum for the differential ac Stark shift~\cite{HayesThesis}, however these Stark shifts must still be taken into account.

Single-qubit gates can be run in a copropagating configuration by applying two rf tones to the same AOM channel, or in a counter-propagating configuration with a separate global addressing beam to support direct $X$ and $Y$ rotations of the qubit state. The global addressing beam is anti-parallel to the IA beams and has a large beam waist to support simultaneous addressing of all qubits.  Two-qubit M\o lmer-S\o rensen~\cite{Molmer1999} gates are run in a counter-propagating configuration, and drive motional excitations via red- and blue-detuned sideband frequencies, requiring two sideband tones as well as a third tone on a counter-propagating beam to complete each Raman process. These gates must be run simultaneously across 3 output channels (two IA beams and the global beam).

The average cycle time is bounded by state preparation and measurement stages, where Doppler-cooling\footnote{Doppler-cooling is the initial step for state preparation and is implemented using a subsystem that is independent from the framework used to drive quantum gates} times are $\approx$ 1 ms, and detection times are $\approx$ 400 $\mu$s.  Single-qubit gates are primarily copropagating as they minimize errors associated with phase uncertainty, yielding higher gate fidelities, and have gate times on the order of 10 $\mu$s. Two-qubit M\o lmer-S\o rensen gate times are typically around 200 $\mu$s. Coherence times ($T_2^*$) exceed 12 s, allowing for potentially large circuit depths.

The control system contains various features specific to the experimental hardware, such as frequency feedforward corrections to account for drift in the pulsed laser cavity length~\cite{Islam2014,Mount2016} and dynamic crosstalk compensation that can support optical crosstalk as well as electronic or acoustic crosstalk from sympathetic coupling in the multichannel AOM\footnote{Cross-talk compensation is more broadly applicable to other QC systems. However, the channel interconnectivity in Octet is set for nearest- and next-nearest-neighbor channels. While connectivity is easily reconfigured in the firmware design, dynamic reconfiguration isn't currently supported.}. However, the most relevant platform-specific details discussed here pertain to requirements needed for coherently switching between different beam configurations for single- and two-qubit Raman gates, and integrated handling of ac Stark shifts induced by the optical transition.

\subsection{Control hardware}

The coherent control system, called ``Octet''~\cite{b1}, uses a custom firmware design implemented on a Xilinx rf system-on-chip (RFSoC). The RFSoC is fitted with programmable logic (PL), a quad-core application processing unit (APU), a dual-core real-time processing unit (RPU), eight 6.554 GSPS digital-to-analog converters (DACs), and eight 4.096 GSPS analog-to-digital converters (ADCs).  Leveraging the hard-core processors limits PL-side resource usage associated with soft-core processors, commonly used in FPGA-based control systems~\cite{artiq}, while maintaining access to real-time functionality with the RPU, and computational power with the APU~\cite{ArtiqSoC,NegnevitskyThesis}.

Although certain features of the Octet design are specific to trapped-ion systems similar to QSCOUT, many features easily translate to other QC platforms.  This includes the ability to perform fast, continuous modulation\footnote{Modulation parameters are updated at a frequency of 409.6 MHz during continuous modulation.} of all waveform parameters (frequency, phase, amplitude, as well as virtual ``frame rotations'' discussed in~\ref{virtualGates}) on all channels simultaneously using on-chip cubic spline interpolators~\cite{Bowler2015}, multi-channel and multi-board synchronization of output waveforms, and integrated gate sequencers for scheduling pulses. 

These more general features share a common theme that embodies one of the underlying design philosophies of Octet: use compact gate representations that remove the need for manual bookkeeping. Manual bookkeeping predominantly relates to phase information, which is handled using low-level firmware design elements in Octet as discussed in section \ref{hwBookkeeping}. Other design elements relate to compact gate representations which are compressed and stored in a series of PL lookup tables (LUTs) detailed in section \ref{hwGateRepresentations}. 

Storing gate data in PL enables parallel readout of gate data, with gates as short as $\approx$ 20 ns, while relaxing requirements for network transfer bandwidth and direct memory access (DMA) throughput.  It also supports low-latency ($\approx$ 50 ns) conditional execution of gate sequences for quantum error correction (QEC), discussed in section \ref{fastBranchingHW}. Compressing gate data additionally simplifies algorithmic sequencing of gates, providing a more tenable framework for interoperability with the RPU. While the RPU is deterministically timed, it is not well suited to computationally intensive tasks. The APU does not support deterministic timing, but can handle more complex algorithmic tasks such as pulse compilation, which requires fitting spline data, custom compression techniques specific to the layout of the PL LUTs, and handling of gate definitions. Gates can be defined off chip and fetched as needed using modern network protocols like Google Remote Procedure Call (gRPC)~\cite{gRPC}. Gates can also be defined on chip, using languages such as Lua~\cite{Lua} or Julia~\cite{Julia-2017} to provide a user-friendly interface which supports dynamic code loading and high-level math functions for calculating modulation parameters.

Leveraging these resources offers a featureful and powerful design, capable of executing gate sequences with deterministic timing, on-chip mutation of gate definitions via high-level algorithms for optimization or autocalibration, including continuous calibration methods to handle slow drift in control parameters. It also works naturally with more resource-intensive off-chip modes of operation such as variational quantum eigensolvers (VQE) or gate definitions which employ physical modeling or machine learning.

While other research efforts have demonstrated algorithmic feedback techniques for mitigating drift on chip~\cite{Blume2017,Negnevitsky2018}, these systems had limited pulse-level control. The flexible multi-parameter spline-based modulation supported by Octet fulfills complex pulse-shaping requirements for a variety of dynamical-decoupling gates and pulse engineering techniques~\cite{Zhu2006,Roos2008,merrill2014,BiercukPMGates,BrownFMGates,Gokhale2019}.  

However, the computational overhead for generating such modulation data is wide-ranging and augmented by implicit requirements for fitting splines and encoding the output for PL. This poses additional challenges for high-level feedback on gate definitions, where affected data may comprise numerous parameters that change non-trivially.
Of course, performance of the embedded software is critical for the realization of such a system. But initial gains originating from features of the hardware design shape the software requirements.

\section{Hardware-native phase bookkeeping}\label{hwBookkeeping}

\subsection{Global phase synchronization}\label{globalPhaseSync}

To simplify pulse-level representations, Octet uses a custom dual-tone direct digital synthesizer (DDS) module to generate sinusoidal baseband waveform data.  This requires much less data than arbitrary waveform generators (AWGs), where waveforms are specified point by point. 

In many cases, gates require different sets of frequencies, such as single-qubit operations and two-qubit M\o lmer-S\o rensen gates. However, changing frequencies nearly always affects the phase relationship between gates run before and after the frequency update, due to the free-running DDS accumulator.  While multiple DDS cores, or even multiple accumulators, can be used to independently track frequencies, this approach quickly becomes tedious, wastes design resources, and may still present challenges when frequency modulated pulses are used.

Rather than track pulse durations and manually override accumulator values with pre-computed phases, Octet dynamically computes phases using a global counter which is multiplied against the frequency word inputs for each DDS.  A DDS accumulator can then be triggered to update its value with this phase. This allows for arbitrary changes to the DDS frequency with the ability to return to the original frequency and phase at a later point in time\footnote{This of course requires that the original frequency had been applied with a synchronization pulse.}, which holds true as long as the board is powered and the global counter isn't reset\footnote{While regularly resetting the global counter is optional, and can be triggered simultaneously with gate sequences, it is unnecessary since absolute phase control is almost never needed with this paradigm. However, absolute phase control can still be achieved by synchronizing against a frequency of 0 Hz and subsequently applying the desired frequency without a synchronization pulse.}. Because the global counter is common to all DDSs, the ability to synchronize across all channels is automatically built in. 

Eliminating the need to track phase information not only cuts down on the amount of unique pulse data, it also sidesteps complications that arise in situations where timing is non-deterministic. While timing determinism can easily be maintained, it often isn't necessary assuming pulses that are run in parallel across different channels remain concurrent, and the introduction of non-deterministic processes doesn't run up against the onset of decoherence.

The ability to introduce non-deterministically timed processes within a circuit opens up a lot of possibilities with respect to hybrid techniques that rely on conventional classical computing resources. For example high-level algorithms can be implemented without being constrained to limitations imposed by real-time processors. It also circumvents challenges that arise when interfacing with other hardware, such as external triggers missing clock edges or operating on different clock domains.

\subsection{Virtual gates}\label{virtualGates}

Additional phase concerns relate to cases in which certain operations are not directly accessible. For example, systems with control fields fixed to a particular axis support $X$ and $Y$ gates, but not $Z$ gates. $R_z(\theta)$ gates, which rotate the qubit state around $\hat{z}$ by an amount $\theta$, can be achieved via basis transforms with sequences of the form $\sqrt{X}^\dagger R_y(\theta)\sqrt{X}$.  However, it is more efficient to virtualize $Z$ gates by adding phase offsets to all subsequent gates, effectively rotating the frame of the qubit. The means in which virtual gates are implemented can ultimately lead to significant overhead for data footprint or compilation time.

Some techniques for virtualizing $Z$ gates employ temporary frequency shifts to advance the DDS phase accumulator~\cite{EfficientZGatesIBM}. While this method allows $Z$ gates to be defined as simple primitives, it presents other challenges when working with Raman gates\footnote{Namely, the effective phase of the drive results from the phase \emph{difference} of the two tones. Phase relationships on the two tones can differ for copropagating single-qubit gates and two-qubit M\o lmer-S\o rensen gates, e.g. when red- and blue- sidebands are applied from an individual addressing channel, in which case both sideband tones require the same virtual phase offset. While this can be avoided by applying sideband tones on the global beam, it limits the beam configurability for two-qubit gates.} or when the DDS frequency needs to change to target another quantum transition. However, This technique is rendered moot with automatic global phase synchronization, as discussed in~\ref{globalPhaseSync}.

An alternative approach is to modify circuits at the software level~\cite{EganFaultTolerant}, but this potentially leads to a larger set of unique gate definitions and suffers from issues with context-dependency. Namely, gate sequences that are conditioned on mid-circuit measurements require branching, that affects the modified sequences downstream if $Z$ gates are used in the branch, leading to multiple variations of any subsequent data. Even if virtual $Z$ gates are avoided during branching, variations may still be required if ac Stark shifts are induced by the conditionally-executed gates.

Instead of manually bookkeeping phase offsets for virtual $Z$ gates, we employ a hardware-based solution to decouple virtual phase, $\phi_z$, from nominal phase, $\phi$. 
\begin{equation}
    \sin(\omega t + \phi) \rightarrow \sin(\omega t + \phi + \phi_z)
\end{equation}
Octet tracks the rotation of the qubit frame\footnote{Virtual $Z$ rotations are often referred to as ``frame rotations'' since the concept is independent of any particular basis.} in dedicated firmware accumulators, which function as persistent phase memories. The accumulator values are treated on the same footing as nominal phase within the DDSs, however the accumulator outputs can be specifically applied to different tones, and can be optionally inverted, to handle variations in beam configurations for single- and two-qubit gates~\cite{InlekPhaseInsensitive}. 

This feature has several advantages. It removes the need for context-dependent representations of subsequent gates, which further reduces the amount of unique data associated with a gate sequence. Moreover, gate sequences that are conditioned on mid-circuit measurements can be {\it temporarily} branched, without needing multiple variations on all data following the conditional sequence, since $Z$ gates can be treated as primitives. 

\subsection{Stark Shifts}\label{acStark}

Another benefit of integrating virtualized $Z$ gates is automatic handling of ac Stark shifts. The energy shift impacts the system in two ways.  First, the change in frequency results in an effective $Z$ rotation. This can be accounted for with a frame rotation that cancels the accumulated phase\footnote{Octet supports the ability to attach a frame rotation to a normal pulse, with the accumulated phase optionally applied before or after the pulse.}.  However, ac Stark shifts pose additional challenges related to synchronization.

Gates must be synchronized to the rotating frame of the bare qubit state, but applied with a frequency tuned to the dressed state. However, this unfortunate and inelegant caveat can be taken care of via continuous modulation of the frame rotation. Namely, applying a frame rotation modulated via a linear ramp generates a constant frequency offset. Laser frequency can be indirectly calibrated with a source that doesn't induce ac Stark shifts (e.g. microwaves).  A measurement of the accumulated phase then determines the overall scale of the modulated frame rotation.

This technique extends to more complex amplitude pulse shaping by specifying the frame rotation with the integral\footnote{Direct integration works in the limit of linear Stark shifts, if the Stark shift is quadratic then additional transformations must be employed.} of the amplitude profile\footnote{The amplitude profile should be transformed in a way that maps onto the desired laser intensity, which for Raman gates is the product of the amplitude profiles of the two tones}. Not only does this remove the phase synchronization caveat in the presence of ac Stark shifts, it offers a simple means of tracking amplitude-dependent frequency shifts while automatically accounting for the accumulated phase needed by subsequent gates.

\section{Gate representations in hardware}\label{hwGateRepresentations}

\subsection{Spline modulation}

All parameters, amplitude, frequency, phase, and frame rotation, support smooth modulation via cubic splines, which are interpolated in PL using a lightweight model that requires only addition operations~\cite{SplineInterpolationNIST,Bowler2015}. Splines offer enormous advantages in terms of reducing data size, since modulation techniques (e.g. dynamical-decoupling gates) generally require orders of magnitude less bandwidth than the waveforms' baseband frequency\footnote{Bandwidth-limited pulse envelopes are often desired to prevent motional-mode excitations, where the lowest mode frequencies for an $^{171}\text{Yb}^+$ system typically exceed 100~kHz, compared to the drive frequencies of roughly 200~MHz for the acousto-optic modulators (AOMs).}. However, synchronous, parallel operation of the all 64 (4 parameters $\times$ 2 tones $\times$ 8 channels) spline engines can still require a significant amount of data.

Coefficients are passed in 256-bit packets, where each coefficient is 40 bits\footnote{40-bit word sizes are used for frequency (giving a resolution of 745~$\mu$Hz), phase, and frame rotation. The same format is used for amplitude, despite its 16-bit word size, to maintain consistency throughout the design.}, giving a combined size of 200 bits for 4 coefficients and a 40-bit duration---duration is needed as a 5\textsuperscript{th} parameter due to the way spline coefficients are transformed to work with the PL-side interpolators---as well as 56 bits of metadata used for routing, pulse attributes like synchronization flags, and other design-specific features. This data size is the same for square pulses, where the higher-order coefficients are set to zero. While this leads to more data overhead, it was a design choice made in order to maintain consistent data flow and reduce complexity in the firmware design until it becomes a dominant limitation.

Because of the independent operation of spline engines, we decompose gate descriptions into ``pulselets'', which comprise one or more words of spline coefficient data, for a particular tone and waveform parameter. The pulselets are ultimately sent to first-in first-out (FIFO) buffers, as shown in Fig.~\ref{FIFOFilling}, which feed the spline engines.  Because data must be sent serially via DMA, and routed to the various spline-engine FIFOs, spline engines are initially halted via a ``wait for trigger'' metadata bit attached to the first word of each pulselet at the beginning of a sequence, and triggered only after all the FIFOs have data.

\begin{figure}
    \centering
    \input{figures/AsymmetricPulses}
    \caption{Example of FIFO filling for a single channel. Each row represents a
    different FIFO, where pulselet data are represented as $P^{T}_{i}$, where
    $T$ represents the target tone (either 0 or 1) the modulation parameter,
    $P$, is one of AMP, FRQ, PHS, FRM for amplitude, frequency, phase, and frame
    rotation, respectively, and $i$ is an index to distinguish unique pulselet
    words. The pulselet words shown are assumed to be part of a larger set of
    data, where pulselet indices might be non-contiguous as they are essentially arbitrary values determined by
    the compiler, and pulselets sharing the same parameter, tone, and index are
    identical. Pulselets are consumed in parallel, at a rate specified by each
    pulselet word's duration, represented by the width of each word.  
    By convention, spline knots are equally distributed across the duration of a pulse, though this is not a firm requirement, and different gates (as represented by the two colors) are assumed to have common boundaries for all pulselets.
    \label{FIFOFilling}}
\end{figure}

FIFOs are continually filled with available pulselet data while a gate sequence runs.  This filling is asymmetric to cut down on data which is constant during a gate, and to support simultaneous modulation on parameters with different numbers of spline knots.  Because data are fed serially but consumed concurrently, constraints are imposed on data ordering in order to avoid blocking conditions that lead to FIFO underflows for gates with large parameter asymmetry.

Throughput requirements for direct streaming of coefficients can exceed theoretical limits on DMA throughput~\cite{XilinxDMA} even for 1~$\mu$s gate times, which is longer than the shortest gate times of 20 ns (typically used for virtual $Z$ gates) currently supported by Octet\footnote{This can in principle be reduced to 2.443~ns or 1/(409.6~MHz), but is restricted to 8 clock cycles in software to avoid potential underflow conditions.}. In most cases, circuits comprise gate sequences with largely redundant pulselet data. Data compression significantly cuts down on throughput requirements.

\subsection{Compression}\label{gateCompression}

By virtue of modularizing gate definitions at the pulse level (via global synchronization and virtual $Z$ gates) we can take advantage of redundant gate calls by storing gate definitions directly in PL. The amount of data that needs to be sent to the device is dramatically reduced for circuits with large depth if they contain a comparatively small number of unique gates.  Pulselet data are stored in a set of LUTs implemented with UltraRAM (URAM)~\cite{XilinxURAM} primitives for fast readout.  Representing gates as a series of LUT addresses gives a compression ratio equal to the address width over the data width, which is $W_{addr}/W_{data} = 12/256 \approx 0.047$ for Octet\footnote{The actual LUT that implements storage at this level takes advantage of maximum filling for 3 URAM primitives, each of which have an address width of 12 bits and a data width of 72 bits. Pulselet data are ultimately stored in word sizes of 216 bits to cut down on the available URAM resources. However, much of the metadata is used for upstream control on the PL side and can be removed for final pulselet representations.}. 

\begin{figure}
    \centering
    \begin{tikzpicture}[scale=1.0]
    \tikzstyle{every node}=[font=\scriptsize]
    \def\height{.8} 
    \def\width{.40}
    \def\LUTItems{{"\text{AMP}^0_0", 
         "\text{AMP}^1_0", 
         "\text{FRQ}^0_0", 
         "\text{FRQ}^1_0",
         "\text{PHS}^0_0",
         "\text{PHS}^1_0",
         "\text{FRM}^0_0",
         "\text{FRM}^1_0",
         "\text{PHS}^1_1",
         "\text{AMP}^0_1", 
         "\text{AMP}^1_1", 
         "\ldots",
         "\text{AMP}^0_n", 
         "\text{AMP}^1_n", 
         "\text{AMP}^1_n", 
         "\text{AMP}^1_n", 
         "\text{AMP}^1_n"}};
    \def\gatewidth{.75*\height}
    \def\figcenter{7*\width}
    \def\gatespacing{3.5*\width}
    \def\goffset{0}
    \def\xcenter{\figcenter-\gatespacing+\goffset}
    \def\xbottom{2.0*\height}
    \draw (\xcenter-.5*\gatewidth,2.0*\height) +(0,0) rectangle ++(\gatewidth,\width);
    \draw (\xcenter,2.0*\height+.5*\width) node{$\sqrt{X}$}; 

    \def\ycenter{\figcenter+\goffset}
    \def\ybottom{2.0*\height}
    \def\xxshift{.15*\width}
    \def\xyshift{.2*\width}
    \draw (\ycenter-.5*\gatewidth,2*\height) +(0,0) rectangle ++(\gatewidth,\width);
    \draw (\ycenter,2*\height+.5*\width) node{$\sqrt{Y}$}; 

    \def\gcenter{\figcenter+\gatespacing+\goffset}
    \def\gbottom{2.0*\height}
    \draw (\gcenter-.5*\gatewidth,2*\height) +(0,0) rectangle ++(\gatewidth,\width);
    \draw (\gcenter,2*\height+.5*\width) node{$G$}; 

    \fill[semitransparent,red!60!black] (0,\height) -- (\xcenter-.5*\gatewidth,\xbottom) -- (\xcenter+.5*\gatewidth,\xbottom) -- (8*\width,\height);
    \fill[semitransparent,green!60!black] (0,\height) -- 
                                    (\ycenter-.5*\gatewidth,\ybottom) -- 
                                    (\ycenter-.1*\gatewidth,\ybottom) -- (5*\width,\height);
    \fill[semitransparent,green!60!black] (6*\width,\height) -- 
                                    (\ycenter+.0*\gatewidth,\ybottom) -- 
                                    (\ycenter+.5*\gatewidth,\ybottom) -- (9*\width,\height);
    \fill[semitransparent,blue!60!black] (2*\width,\height) -- 
                                    (\gcenter-.5*\gatewidth,\ybottom) -- 
                                    (\gcenter-.2*\gatewidth,\ybottom) -- (6*\width,\height);
    \fill[semitransparent,blue!60!black] (9*\width,\height) -- 
                                    (\gcenter-.1*\gatewidth,\ybottom) -- 
                                    (\gcenter+.5*\gatewidth,\ybottom) -- (14*\width,\height);
    \foreach \x in {0,...,13}
        {
            \pgfmathsetmacro{\n}{\LUTItems[\x]};
            \draw (\x*\width,0) +(0.,0.) rectangle ++(\width,\height);
            \draw (\x*\width+.5*\width,.5*\height) node[rotate=-90]{$\n$};
        }
\end{tikzpicture}
    \caption{Shared pulselet data for different gates. In this case $\sqrt{X}$
    and $\sqrt{Y}$ gates have a large overlap of identical pulselet data with
    the exception of a phase word. Other gates (such as a hypothetical $G$ gate)
    might also share the same pulselet data, meaning a gate's pulselet data
    might be arbitrarily ordered in this representation.}
    \label{SharedGateData}
\end{figure}
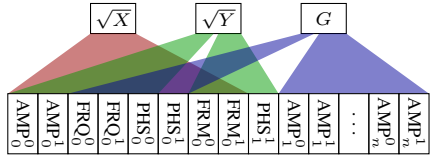

Due to the limited number of URAM primitives available on the device, efficient representation of gate data is imperative. We further compress pulselet data which are shared among multiple gates, thus distilling gates into a minimal set of unique pulselet data as shown in Fig.~\ref{SharedGateData}. This can have a significant impact on memory footprint as certain classes of gates are often nearly identical, e.g. $X$ and $Y$ gates only differ by phase. In other cases, they can be defined in ways that maximize the overlap with other gates, such as matching gate times and implementing gates such as $X$ and $\sqrt{X}$ with different amplitudes as opposed to different durations.  


Additional compression of pulselet data involves a custom scheme specific to the LUT topology implemented in the PL ``gate sequencer modules''. The LUTs, in conjunction with other firmware elements, are configured as a multi-stage decompression pipeline in order to reconstruct the original gate. Because a gate boils down to a list of addresses of locally-stored pulselet data, more efficient representations could be achieved if these addresses were contiguous.  In this case, a gate could be represented simply by its address bounds, which could be iterated over directly in PL. However, sharing of unique pulselet data makes this practically impossible, so a ``Mapping LUT'' (MLUT) is introduced to remap pulselet addresses, for the ``Pulse LUT'' (PLUT), into a contiguous format\footnote{The MLUT uses 14-bit addresses, compared to the 12-bit addresses of the PLUT, to take advantage of gains from shared data where more redundancy is needed. It is implemented with a single URAM primitive, but uses a byte-write scheme to pack multiple PLUT addresses into each 72-bit entry to achieve the larger address width.}. The MLUT is paired with an iterator module to step through input memory bounds.

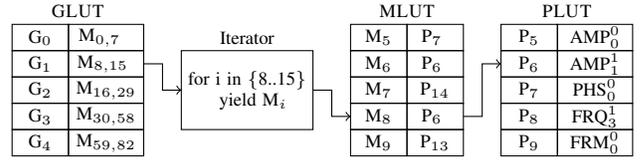
\begin{figure}
    \centering
    \begin{tikzpicture}[scale=1.0]
    \tikzstyle{every node}=[font=\scriptsize]
    \def\height{1.75} 
    \def\widths{{.75, 1.0, .5,  
                 .75, 1.0, .5,  
                 .75, .75, .5,  
                 .75, 1.0,  0}} 
    \def\LUTItems{{
        {"\text{G}_0", "\text{G}_1", "\text{G}_2", "\text{G}_3", "\text{G}_4",}, 
        {"\text{M}_{0,7~~}", "\text{M}_{8,15~}", "\text{M}_{16,29}", "\text{M}_{30,58}", "\text{M}_{59,82}"}, 
        {}, {}, {}, {},
        {"\text{M}_5", "\text{M}_6", "\text{M}_7", "\text{M}_8", "\text{M}_9", }, 
        {"\text{P}_{7~}", "\text{P}_{6~}", "\text{P}_{14}", "\text{P}_{6~}", "\text{P}_{13}", }, 
        {},
        {"\text{P}_5", "\text{P}_6", "\text{P}_7", "\text{P}_8", "\text{P}_9", }, 
        {"\text{AMP}^0_0", "\text{AMP}^1_1", "\text{PHS}^0_0", "\text{FRQ}^1_3", "\text{FRM}^0_0", }, 
    }}
    \pgfmathsetmacro{\itstart}{\widths[0]+\widths[1]+\widths[2]};
    \pgfmathsetmacro{\itwid}{\widths[3]+\widths[4]};
    \pgfmathsetmacro{\itwidtxt}{0.5*\itwid};
    \draw (\itstart,0) +(0.,-.2*\height) rectangle ++(\itwid,-1.*\height*.2*4);
    \draw (\itstart+\itwidtxt,-1.*\height*.1*5) node[text width=2cm,align=center]{for i in \{8..15\}\\~~yield $\text{M}_i$};
    \pgfmathsetmacro{\myt}{0}
    \foreach \x/\tag/\wid/\gap in {0/G/0.75/0,1/M/1.75/2.75, 6/M/.75/0,7/P/0.75/.5, 9/P/0.75/0,10/R/1.75/0}
    {
        \pgfmathsetmacro{\widt}{\widths[\x]}
        \pgfmathsetmacro{\tot}{\myt}
        \foreach \y in {0,...,4}
        {
            \pgfmathsetmacro{\n}{\LUTItems[\x][\y]};
            \draw (\tot,-.2*\y*\height) +(0.,0.) rectangle ++(\widt,-.2*\height);
            \draw (\tot+0.5*\widt,-.2*\y*\height-.1*\height) node{$\n$};
        }
        \global\let\myt=\tot;
        \pgfmathsetmacro{\tot}{\myt+\widt+\gap}
        \global\let\myt=\tot;
    }
    \pgfmathsetmacro{\starta}{\widths[0]+\widths[1]};
    \pgfmathsetmacro{\da}{\widths[2]};
    \pgfmathsetmacro{\startb}{\starta+\da+\widths[3]+\widths[4]};
    \pgfmathsetmacro{\db}{\widths[5]};
    \pgfmathsetmacro{\startc}{\startb+\db+\widths[6]+\widths[7]};
    \pgfmathsetmacro{\dc}{\widths[8]};
    \pgfmathsetmacro{\wida}{\widths[0]+\widths[1]};
    \pgfmathsetmacro{\widb}{\widths[3]+\widths[4]};
    \pgfmathsetmacro{\widc}{\widths[6]+\widths[7]};
    \pgfmathsetmacro{\widd}{\widths[9]+\widths[10]};
    \draw[->] (\starta,-.2*1.5*\height) -- ++(.5*\da,0) -- ++(0,-.2*\height) -- ++(0.5*\da,0);
    \draw[->] (\startb,-.2*2.5*\height) -- ++(.5*\db,0) -- ++(0,-.2*\height) -- ++(0.5*\db,0);
    \draw[->] (\startc,-.2*3.5*\height) -- ++(.5*\dc,0) -- ++(0,.4*\height) -- ++(0.5*\dc,0);
    \draw (\starta-.5*\wida,.1*\height) node[text width=2cm,align=center]{\textsc{GLUT}};
    \draw (\startb-.5*\widb,-.1*\height) node[text width=2cm,align=center]{\text{Iterator}};
    \draw (\startc-.5*\widc,.1*\height) node[text width=2cm,align=center]{\textsc{MLUT}};
    \draw (\startc+\dc+.5*\widd,.1*\height) node[text width=2cm,align=center]{\textsc{PLUT}};
\end{tikzpicture}
    \caption{Gate sequencer module. Each gate sequencer reads in an 11-bit
    ``gate identifer'', which is used as an address in the GLUT. The GLUT data
    comprise two 14-bit address bounds, which are iterated over and passed to
    the MLUT. The MLUT is simply used to remap the 12-bit PLUT addresses to
    allow for specific ordering when read out by the iterator module, and can
    also store a larger number of addresses than the PLUT to account for data
    redundancy. The PLUT contains the lowest-level pulselet data used to feed
    the spline engine FIFOs.
    }
    \label{gateSequencerFig}
\end{figure}

A final compression stage is added with a ``Gate LUT'' (GLUT), where the full configuration is shown in Fig.~\ref{gateSequencerFig}. The GLUT simply stores the memory bounds for a gate, which is then indexed by the GLUT address. The improvement in the compression factor, which is $11/28\approx0.4$ for current system parameters, is somewhat marginal at this point\footnote{GLUT address widths are 11 bits, which can support the maximum number of non-overlapping gates in the MLUT, since 8 words are needed at a minimum for a gate. If requirements allow for fewer unique gates in general, the GLUT address could be reduced to give a larger compression ratio. However, the number of gates stored in the GLUT can be much larger than the number of unique gates applied on a channel in order to account for wait times during the execution of gates on other channels.}.  However, its primary benefit is that it enables fast branching in gate sequences.

\subsection{Fast branching}\label{fastBranchingHW}

Short gate sequences conditioned on mid-circuit measurements can be handled entirely in PL if all possible sequences are known in advance. This technique uses hybrid GLUT addresses formed by a combination of gate sequence bytecode and external hardware inputs. This technique is ``fast'' in the sense that the latency ($\approx$ 50 ns) is dominated by the time it takes for a gate sequence to propagate through the LUTs. The bytecode, which contains multiple densely-packed GLUT addresses, also contains metadata that indicates whether the current packet should be interpreted as a normal gate sequence or a ``branch sequence'', which depends on a mid-circuit measurement. 

For a new mid-circuit measurement, the gate sequencer is temporarily halted until a measurement result is obtained. If a branch sequence exceeds the size of a single packet, a slight variation in the metadata can indicate a continuation of a branch sequence. The packed GLUT addresses in a branch sequence are modified via a bitwise \texttt{or} with registers containing the measurement result, namely 
\begin{equation}
A^{\textsc{glut}}_n | (O^{\textsc{meas}} \ll S),
\end{equation}
where bitwise operators retain their meaning from the C programming language, $A^{\textsc{glut}}_n$ is the n$^\text{th}$ GLUT address in the sequence bytecode, $O^\textsc{meas}$ is the binary measurement outcome, and $S$ is a configurable shift used to constrain measurement inputs to a certain range (if desired).

By convention, an additional term is included\footnote{This additional bit can be configured via a PL register that acts as a ``soft'' measurement input. The MSB in this case applies to the full 12 bit GLUT address used for programming---not the 11-bit GLUT address used for streaming---to give a clear separation between gates in each case. However, the PL register used for the MSB can simply be an extension of the measurement data if one chooses, but this will likely complicate compilation.} in the \texttt{or} operation, giving 
\begin{equation}
A^{\textsc{glut}}_n | (O^{\textsc{meas}} \ll S) | (1\ll(W^\textsc{glut}_{addr}-1)),
\end{equation}
where $W^\textsc{glut}_{addr}$ is the GLUT address width.  This sets the most significant bit (MSB) of the GLUT address to one in order to distinguish between gates in normal sequences and branch sequences when the measurement outcome is zero.

\section{Pulse compiler}

\subsection{Parser}

We use Jaqal~\cite{Landahl2020,Morrison2020} as our target quantum assembly language. The official Jaqal parser from the JaqalPaq package~\cite{jaqalpaq} is written in Python using the sly parser generator package. However, for a more uniform software framework that offers higher performance (especially on an embedded system) we implement a new parser written in Go~\cite{golang}. The parser has three main components: a tokenizer, a recursive descent parser, and a semantic analyzer. Further processing is accomplished using a visitor design pattern~\cite{GangOfFour}.

The tokenizer (also called a lexer) is a standard, but not universal, stage used to convert a stream of characters into tokens such as keywords, identifiers, bracketing characters, and so on. The tokenizer is able to take advantage of Jaqal's lack of unicode support and small lookahead requirements. By reusing the memory allocated for previously-returned tokens, dynamic memory allocation is eliminated, but the advantages of passing around pointers are maintained.

The Jaqal grammar is well-suited to a recursive descent parser. The main weakness of recursive descent parsers is their inability to handle left-branching grammar rules~\cite{Holub90}. Since Jaqal lacks such rules, due largely to its lack of arithmetic operations, special techniques are not required to parse it.  The parsing time is thus dominated by the combination of the tokenizer and dynamic memory allocation of the intermediate representation (IR).

After parsing, semantic analysis is performed on the resulting IR. This step determines what each identifier is referencing. An identifier may refer to a constant\footnote{Constants, defined via ``let'' statements, are immutable within a Jaqal program; however, they are often treated as variables by Octet allowing a Jaqal file to effectively be parameterized by these constants.}, a register, a gate name, a macro, a macro parameter, or a named qubit, and differentiating between those based on context is useful for other program transformations.  To avoid additional dynamic memory allocations, these objects are reused whenever possible. However, in order to prevent hard-to-debug errors, all IR objects are treated as immutable once created. 

The output of the parser is a tree IR that reflects the hierarchical structure of Jaqal. But another form is sometimes more convenient. We call this form the Tabulated Intermediate Representation (TIR). This form uses tables for gates, blocks, and macros. Each table entry is given a globally unique index. Gate entries record the name of the gate and its arguments, block entries record the type of block and the statements (gates, macro definitions, or other blocks) it contains, and the macro table contains macro definitions. Importantly, entries in the gate table which would compile to the same pulse sequence are combined into the same entry. In practice, all gates contained in parallel blocks are considered to have possibly unique pulse sequences (due to duration matching of concurrent gates, discussed in section \ref{MergingAndNOPPadding}), and gates not contained within parallel blocks are considered to have the same pulse sequence if they have the same gate name and argument values. 

The TIR thus provides substantial compression in the common case where gates are repeated. This representation also reduces the asymptotic complexity of certain transformations such as macro substitution. Rather than traversing the entire circuit, transformations may act directly on the gate table, which may be significantly smaller.

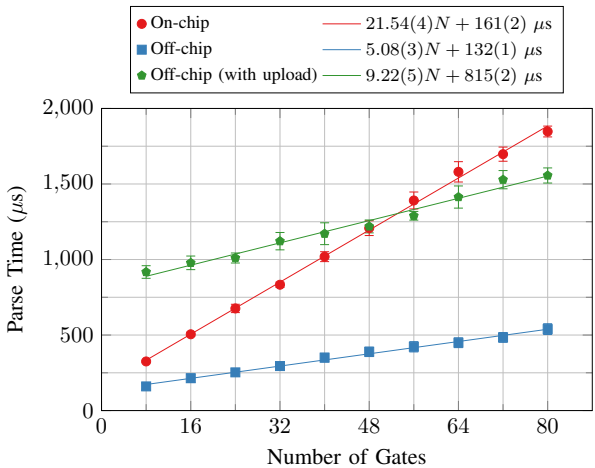
\begin{figure}
    \centering
    \scalebox{.85}{
    \begin{minipage}[t]{.85\textwidth}
        \vspace{0pt}
        \definecolor{d1}{rgb}{0.8941176470588236,0.10196078431372549,0.10980392156862745}
\definecolor{d2}{rgb}{0.21568627450980393,0.49411764705882355,0.7215686274509804}
\definecolor{d3}{rgb}{0.20196078431372547,0.5862745098039216,0.1901960784313726}
\definecolor{d4}{rgb}{0.85,0.3980392156862745,0.0}
\definecolor{d5}{rgb}{0.596078431372549,0.3058823529411765,0.6392156862745098}
\begin{tikzpicture}
    \pgfplotsset{    
    every axis x label/.style={
        at={(ticklabel* cs:1.05)},
        anchor=west,
    },
    every axis y label/.style={at={(current axis.above origin)},
    anchor=north east,
    yshift=1cm,},
    compat=1.17,
    every axis legend/.append style={
        legend columns=3,font=\footnotesize},
        width=\axisdefaultwidth*1.1,
    }
    \begin{axis}[
      cycle multiindex* list={
        d1,d2,d3\nextlist
        solid,{solid, mark options={solid}}\nextlist
        mark=*,mark=square*,mark=pentagon*,mark=triangle*,mark=diamond*
      },
      xtick={0,16,32,48,...,80},
      minor tick num=1,
      width=\axisdefaultwidth*0.91,
      height=\axisdefaultheight*.65,
      xmin=0,xmax=88,
      ymin=0,ymax=2000,
      xlabel={Number of Gates},
      ylabel={Parse Time ($\mu$s)},
      grid=both,
      legend cell align=left,
      transpose legend,
      legend style={at={(0.5,1.05)},anchor=south},
      scale only axis,
    ]
      \addplot+ [only marks, error bars/.cd, y dir = both, y explicit] 
                table [col sep=comma,header=false,x index=0,y index=1,y error index=2] 
                    {figures/poseallfullydat.csv};
      \addplot+ [only marks, error bars/.cd, y dir = both, y explicit] 
                table [col sep=comma,header=false,x index=0,y index=1,y error index=2] 
                    {figures/posallfullydat.csv};
      \addplot+ [only marks, error bars/.cd, y dir = both, y explicit] 
                table [col sep=comma,header=false,x index=0,y index=1,y error index=2] 
                    {figures/posuallfullydat.csv};
      \addplot+ [mark=none,domain=8:80]{21.54*x+161.0};
      \addplot+ [mark=none,domain=8:80]{5.08*x+132.0};
      \addplot+ [mark=none,domain=8:80]{9.22*x+815.0};
      \legend{On-chip,
              Off-chip,
              Off-chip (with upload),
              $21.54(4)N + 161(2)~\mu$s,
              $5.08(3) N + 132(1)~\mu$s,
              $9.22(5) N + 815(2)~\mu$s,
              }
    \end{axis}
\end{tikzpicture}
    \end{minipage}
    }
    \caption{Comparison of on- and off-chip parse times. Parsing is $\approx$ 4
    times faster off chip. However, parsing on chip is more favorable for
    circuits with fewer than 50 gates due to the additional upload overhead. Less than 0.1\% of the off-chip upload results exceed 5$\sigma$ (up to 40 ms) and are not included.}
    \label{parseTimeBenchmark}
\end{figure}

Parsing can either be performed on- or off-chip, depending on use case.  Performance is compared for on- and off-chip parsing in Fig.~\ref{parseTimeBenchmark}. While the off-chip parsing clearly outpaces on-chip parsing, this is offset by upload times, in which case it is more favorable to parse short circuits ($\lesssim$ 50 gates) on chip.

\subsection{Gate definition interface}

Off-chip gate definitions use ``JaqalPaw''~\cite{JaqalPaw}, the Python-based software developed for pulse-level gate definitions on Octet.  On-chip gate definitions use comparable conventions, as they pertain to Octet hardware features, and to simplify translation to gates described in other programming languages as discussed in section~\ref{gateFetching}.

For systems like QSCOUT, gates require pulses that are executed simultaneously on one or more channels, and in some cases might employ several concatenated pulses where pulselet data is logically separated between pulses. Gate data are thus represented as an array of independent pulse objects, which in turn describe lower-level pulselet data.  Each pulse must, at a minimum, specify the target channel and duration, where all pulselet data and metadata defaults to zero. Pulses which \emph{only} specify channel and duration are effectively $I$ gates, with a certain wait time, and are called NOPs when referring to pulse-level gate descriptions.

Within each pulse, pulselets are either scalar values or array types that distinguish between spline knots and discrete updates (for which higher-order coefficients are set to zero), which can be combined as a mixed type. Splines and discrete arrays are equally distributed over the pulse duration.  While equal time distribution is not a strict hardware requirement, it simplifies pulse-level descriptions and is sufficient in most cases. Mixed types, which are arrays of scalar, discrete, spline, or additional mixed types, allow pulses to chain together different modulation types. When nested, they represent a tree-like structure, where each top-level element is equally distributed over the duration of the pulse, and nested structures are distributed over the subdivided durations.

By convention, pulses on different channels are aligned with the beginning of a gate, and multiple pulses on the same channel are run back-to-back. This allows more control of piecewise modulation that requires non-uniform time distribution or variations in pulse metadata. If different pulse alignment is required, it can be achieved using NOPs with the desired duration.

Pulse metadata, used to control hardware-level operations such as enabling frequency feedforward corrections or synchronization flags, are associated with the entire pulse. Metadata is typically set for each word in the low-level pulselet data, such as frequency feedforward enables, however certain types of metadata are applied only to the first pulselet word, such as synchronization flags or metadata used to wait for an external trigger.

\subsection{Gate fetching}\label{gateFetching}

The first step of the compilation after parsing is to collect pulse-level gate definitions. Unique gate calls are determined by a combination of the gate's name and input arguments, and are read out from the gate table in the TIR, where redundant gate calls have already been removed by the parser as a precompression step. Gate names are used to determine if the gate is defined on or off chip and subsequently retrieved from the appropriate source. 

On-chip generation of gate definitions can in principle use a variety of interfaces. While gate definitions can be directly implemented in Go, this can lead to a certain degree of tedium in terms of cross compiling, restarting binaries, and a less user-friendly interface. Go supports embedding of C code (via ``cgo'') and, likewise, code which can be embedded in C. Candidate languages we have tested with this approach are Julia~\cite{Julia-2017}, Lua~\cite{Lua}, and LuaJIT~\cite{LuaJIT}.  While Lua and LuaJIT work well with direct embedding via cgo, Julia poses certain thread-safety challenges when directly embedded in Go. To circumvent these issues, we instead run Julia as a separate process and use a combination of memory maps and semaphores for inter-process communication (IPC).  Because of the broad applicability of memory maps and semaphores, this approach can be extended to a number of languages. However, because of the slight speed advantage and large number of natively-supported math operations, we are currently focusing efforts on a Julia-based API for on-chip gate definitions.

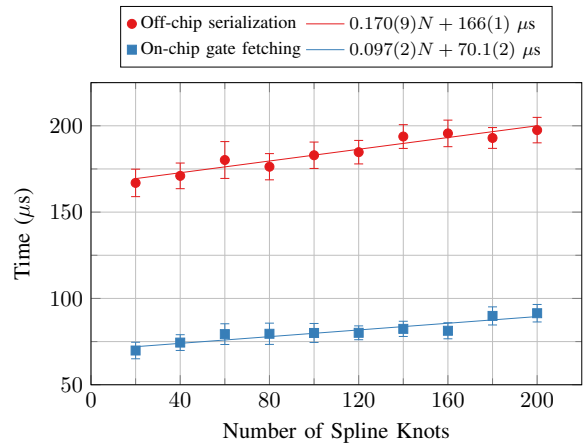
\begin{figure}
    \centering
    \scalebox{.85}{
    \begin{minipage}[t]{.85\textwidth}
        \vspace{0pt}
        \definecolor{d1}{rgb}{0.8941176470588236,0.10196078431372549,0.10980392156862745}
\definecolor{d2}{rgb}{0.21568627450980393,0.49411764705882355,0.7215686274509804}
\definecolor{d3}{rgb}{0.30196078431372547,0.6862745098039216,0.2901960784313726}
\definecolor{d4}{rgb}{0.95,0.4980392156862745,0.0}
\definecolor{d5}{rgb}{0.596078431372549,0.3058823529411765,0.6392156862745098}
\begin{tikzpicture}
    \pgfplotsset{    
    every axis x label/.style={
        at={(ticklabel* cs:1.05)},
        anchor=west,
    },
    every axis y label/.style={at={(current axis.above origin)},
    anchor=north east,
    yshift=1cm,},
    compat=1.17,
    every axis legend/.append style={
        legend columns=2,font=\footnotesize},
        width=\axisdefaultwidth*1.1,
    }
    \begin{axis}[
      cycle multiindex* list={
        d1,d2\nextlist
        solid,{solid, mark options={solid}}\nextlist
        mark=*,mark=square*,mark=pentagon*,mark=triangle*,mark=diamond*
      },
      xtick={0,40,80,...,200},
      minor tick num=1,
      width=\axisdefaultwidth*0.91,
      height=\axisdefaultheight*.65,
      xmin=0,
      ymin=50,ymax=225,
      xlabel={Number of Spline Knots},
      ylabel={Time ($\mu$s)},
      grid=both,
      legend cell align=left,
      transpose legend,
      legend style={at={(0.5,1.05)},anchor=south},
      scale only axis,
    ]
      \addplot+ [only marks, error bars/.cd, y dir = both, y explicit] 
                table [col sep=comma,header=false,x index=0,y index=1,y error index=2] 
                    {figures/gnupythonserdata.csv};
      \addplot+ [only marks, error bars/.cd, y dir = both, y explicit] 
                table [col sep=comma,header=false,x index=0,y index=1,y error index=2] 
                    {figures/gnuftallfullcsv.csv};
      \addplot+ [mark=none,domain=20:200]{0.170*x + 166.0};
      \addplot+ [mark=none,domain=20:200]{0.097*x + 70.1};
      \legend{Off-chip serialization,
              On-chip gate fetching,
              $0.170(9)N + 166(1)~\mu$s,
              $0.097(2) N + 70.1(2)~\mu$s,
              }
    \end{axis}
\end{tikzpicture}
    \end{minipage}
    }
    \caption{Times for on-chip fetching vs off-chip serialization of gate
    definitions. In both cases, identical sets of gate definitions are
    calculated.  On-chip fetching is implemented with Julia, and includes the
    time to convert data transferred via memory map to a native Go struct.
    Off-chip gate calculations are performed in Python using JaqalPaw, and
    protobuf serialization is implemented via C++ reflection.
    }
    \label{fetchVsSerializeBenchmark}
\end{figure}

\begin{figure}
    \centering
    \includegraphics[width=80mm]{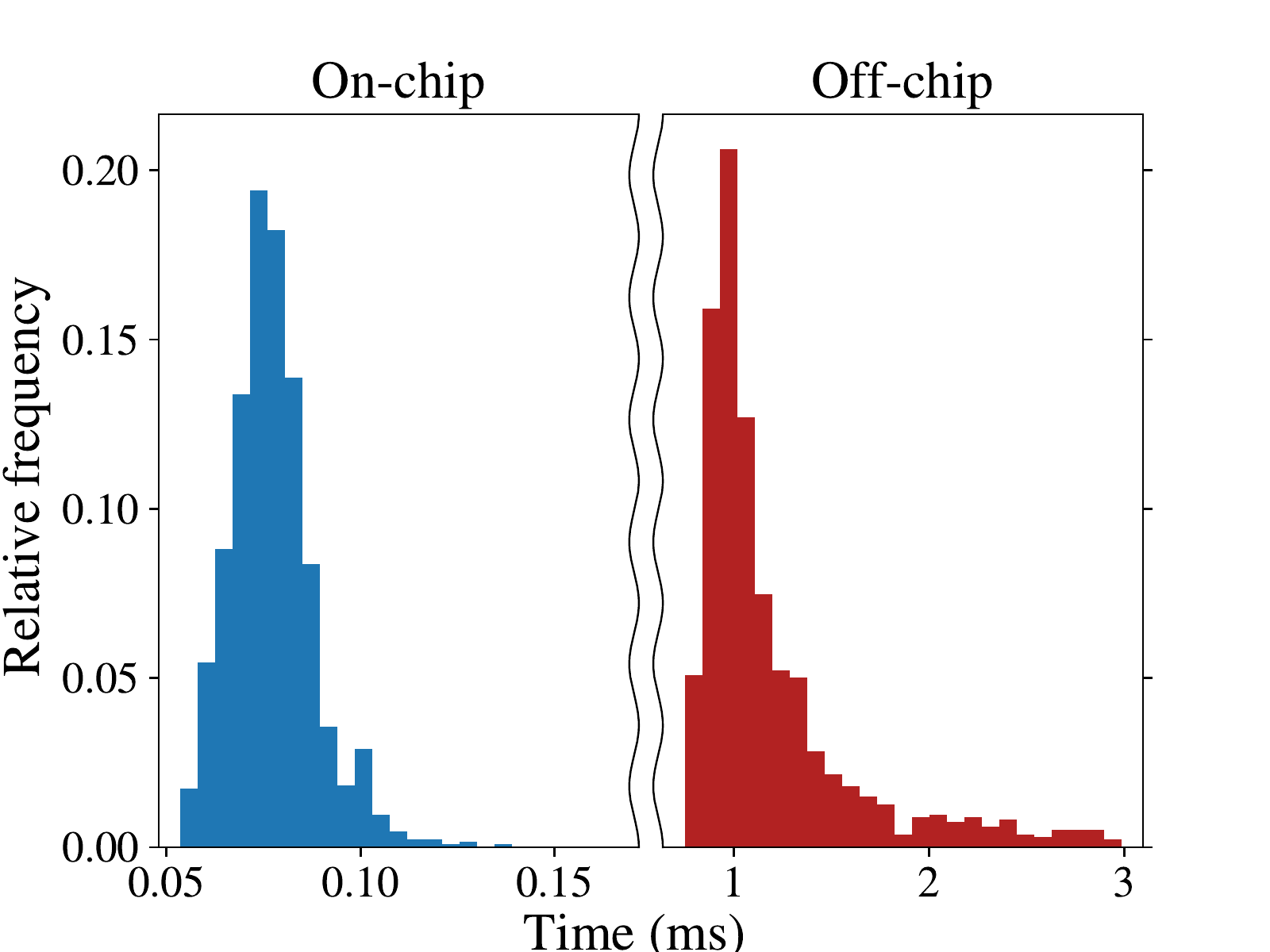}
    \caption{Comparison of on- and off-chip gate fetching. Data is shown as the
    relative number of points that fall within a histogram bin for gates with
    100, 120, and 140 knots.  Both on-chip (left) and off-chip (right) data
    correspond to conditions shown in Fig.~\ref{fetchVsSerializeBenchmark}.
    Off-chip fetch times include a round-trip transaction from a request
    originating on chip, with data returned in a serialized protobuf message
    over gRPC. Less than 0.1\% of the off-chip fetch times are not shown, as they include points with spuriously large transfer times as high as 40 ms.}
    \label{fetchVsTransferBenchmark}
\end{figure}

Gates defined off chip are retrieved via gRPC~\cite{gRPC} over a local network.  The off-chip definitions are serialized using protocol buffers (protobuf), the message format used by gRPC, and sent back to the chip. Differences in the times for on-chip and off-chip fetching are substantial, where protobuf serialization alone exceeds the on-chip fetch times by more than a factor of two, as shown in Fig.~\ref{fetchVsSerializeBenchmark}. When taking network transfer times into account, the times differ by over an order of magnitude as shown in Fig.~\ref{fetchVsTransferBenchmark}.

\subsection{Caching}

In order to prevent redundant calculation and storage of pulselet data, we use memoization at several points throughout the code. The first stage of memoization happens after fetching a gate definition. This definition is reused when compiling subsequent circuits unless gates are specifically invalidated.  Gates can be invalidated on a per-gate basis, or all at once to force a recalculation of all gates. 

Such invalidation is used between circuits where changes to calibration data are anticipated or between circuits which use an entirely different set of gates---for instance, some circuits may be designed to characterize noise in simple gate definitions while others may opt for higher-fidelity dynamical-decoupling gates. However, if one expects gate definitions to remain constant, such as during a VQE algorithm, broad invalidation is not used as it will lead to unnecessary overhead.

The most critical application for caching is for eliminating duplicate data to minimize the amount of pulselet data stored in the PLUT. However, for more efficient handling of pulse-level representations, gates are broken up into simple abstractions for pulses and their constituent pulselets, and organized using ``pulse managers''.

\subsection{Pulse managers}

The compiler is fitted with an array of ``pulse managers'' that are tied to specific channels, since hardware channels each have a dedicated gate sequencer.  The pulse managers are structured to mirror the layout of the gate sequencer LUTs, simplifying the process of packing them before or while running a circuit.  Pulse managers provide easy and efficient access to pulses and their constituent pulselets, allowing higher software levels to mutate pulses at a granular level \emph{in-situ}. Pulses may also share pulselets, so that changes to calibration data may be efficiently applied to a single pulselet, thereby affecting an arbitrary number of gates relying on that pulselet.  

Each pulse manager uses an array of structs to store the pulses and pulselets.  Using a contiguous array rather than a map requires periodic reallocation and copying of data, but has many performance advantages. Each pulse and pulselet is associated with an integer index, allowing efficient lookup. Additionally, since all entries are contiguous, a common case where adjacent pulselets belong to the same pulse takes advantage of the CPUs caches. We also trade a large number of small dynamic memory allocations for a small number of large ones.  Since dynamic memory allocations are time-consuming, and we have strict time requirements, we can over-allocate memory when compiling a Jaqal file, and thereby avoid additional allocations when updating gate and pulse definitions.

\subsection{Merging parallel gates and NOP padding}\label{MergingAndNOPPadding}

Gate scheduling requires continuous filling of the spline FIFOs to preserve relative gate alignment. As a result, any gates which are run sequentially must be padded with NOPs on unused channels. This is essentially the same as sending timer data to each spline FIFO, that triggers a read on subsequent pulselet data after the timer elapses. While these timers could be amalgamated for back-to-back NOP pulses, we choose to split up timer data by aligning NOPs to pulse boundaries in order to maintain a modular, context-free representation.

Gate sequences are thus divided at pulse boundaries, as indicated by the two regions in Fig.~\ref{FIFOFilling}. This can be abstracted across channels using a ``gate slice'' representation, and used to distinguish between time-separated blocks at the quantum assembly level. Each gate slice represents a sequential operation at the top-level of the circuit, which is either an individual gate or multiple gates run in parallel.

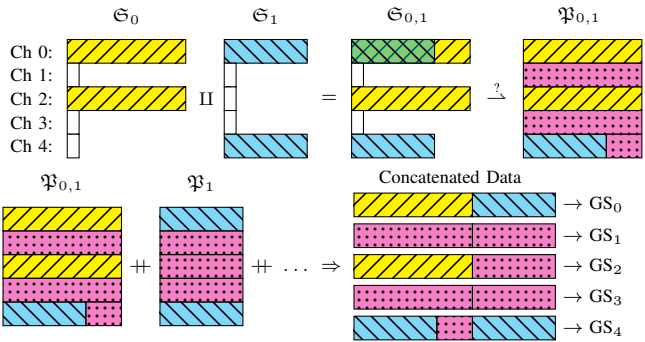
\begin{figure}
    \centering
    \begin{tikzpicture}[scale=0.9]
    \tikzstyle{every node}=[font=\scriptsize]
    \def\height{1.75} 
    \def\width{1.75} 
    \def\offsetscale{1.2}
    \def\shallowwidth{.7}
    \def\fullwidth{1}
    \def\coloone{"semitransparent,white!100!black"}
    \def\colotwo{"semitransparent,blue!60!black"}
    \def\colothree{"semitransparent,red!60!black"}
    \def\colofour{"semitransparent,green!60!black"}
    \def\colofive{"semitransparent,yellow!80!black"}
    \def\ColorTable{{"white!100!black",
                     "blue!60!black",
                     "yellow!100!black",
                     "cyan!100!white",
                     "magenta!100!black"}}
    \def\TransparencyTable{{"semitransparent",
                     "semitransparent",
                    "opaque",
                     "semitransparent",
                     "semitransparent",
                     "semitransparent",
                     "opaque"}}
    \def\PatternTable{{"north eastish lines",
                     "north westish lines",
                     "north westish lines",
                     "north eastish lines",
                     "dots",
                     "hatchish lines",
                     "north west lines, pattern color=black",
                     "north west lines, pattern color=black",
                     "north west lines, pattern color=black",
                     "north west lines, pattern color=black",
                     "semitransparent,yellow!80!black"}}
    \def\PatternColorTable{{"white",
                     "black",
                     "black",
                     "black",
                     "black",
                     "yellow!80!black"}}
    \def\GateInit{{0/1,1/.7,2/.1,3/.1,4/.1,5/.4,6/.1,7/.1}}
    \def\GatePad{{0/1/0,
                  1/.7/0,1/.3/.7,
                  2/1/0,
                  3/1/0,
                  4/1/0,
                  5/.4/0,5/.6/.4,
                  6/1/0,
                  7/1/0}}
    \foreach \y/\x/\c in {0/1/2,
                       1/.1/0,
                       2/1/2,
                       3/.1/0,
                       4/.1/0}
    {
        \pgfmathsetmacro{\colo}{\ColorTable[\c]};
        \pgfmathsetmacro{\pat}{\PatternTable[\c]};
        \pgfmathsetmacro{\opacity}{\TransparencyTable[\c]};
        \pgfmathsetmacro{\patcolo}{\PatternColorTable[\c]};
        \fill[\opacity,\colo] (0*\offsetscale*\width,-.2*\y*\height) +(0.,0.) rectangle ++(\x*\width,-.2*\height);
        \fill[pattern=\pat, pattern color=\patcolo] (0*\offsetscale*\width,-.2*\y*\height) +(0.,0.) rectangle ++(\x*\width,-.2*\height);
        \draw (0*\offsetscale*\width,-.2*\y*\height) +(0.,0.) rectangle ++(\x*\width,-.2*\height);
    }
    \pgfmathsetmacro{\woffset}{.5*(\fullwidth-\shallowwidth+\offsetscale-1)*\width};
    \draw (1*\offsetscale*\width-.1*\woffset,-.2*2.5*\height) node{$\amalg$};

    \foreach \y/\x/\c in {0/.7/3,
                       1/.1/0,
                       2/.1/0,
                       3/.1/0,
                       4/.7/3}
    {
        \pgfmathsetmacro{\colo}{\ColorTable[\c]};
        \pgfmathsetmacro{\opacity}{\TransparencyTable[\c]};
        \pgfmathsetmacro{\pat}{\PatternTable[\c]};
        \pgfmathsetmacro{\patcolo}{\PatternColorTable[\c]};
        \fill[\opacity,\colo] (1*\offsetscale*\width+.5*\woffset,-.2*\y*\height) +(0.,0.) rectangle ++(\x*\width,-.2*\height);
        \fill[pattern=\pat, pattern color=\patcolo] (1*\offsetscale*\width+.5*\woffset,-.2*\y*\height) +(0.,0.) rectangle ++(\x*\width,-.2*\height);
        \draw (1*\offsetscale*\width+.5*\woffset,-.2*\y*\height) +(0.,0.) rectangle ++(\x*\width,-.2*\height);
    }
    \draw (2*\offsetscale*\width-.75*\woffset,-.2*2.5*\height) node{$=$};

    \foreach \y/\x/\c in {
                       0/1/2,0/.7/3,
                       1/.1/0,
                       2/1/2,
                       3/.1/0,
                       4/.7/3}
    {
        \pgfmathsetmacro{\colo}{\ColorTable[\c]};
        \pgfmathsetmacro{\pat}{\PatternTable[\c]};
        \pgfmathsetmacro{\opacity}{\TransparencyTable[\c]};
        \pgfmathsetmacro{\patcolo}{\PatternColorTable[\c]};
        \fill[\opacity,\colo] (2*\offsetscale*\width+0*\woffset,-.2*\y*\height) +(0.,0.) rectangle ++(\x*\width,-.2*\height);
        \fill[pattern=\pat, pattern color=\patcolo] (2*\offsetscale*\width+0*\woffset,-.2*\y*\height) +(0.,0.) rectangle ++(\x*\width,-.2*\height);
        \draw (2*\offsetscale*\width+0*\woffset,-.2*\y*\height) +(0.,0.) rectangle ++(\x*\width,-.2*\height);
    }
    \draw (2*\offsetscale*\width+0.1*\woffset+\offsetscale*\width,-.2*2.5*\height) node{$\displaystyle\pad_{\text{~}}^{\text{?}}$};

    \foreach \y/\x/\o/\c in {0/1/0/2,
                       2/1/0/2,
                       4/.7/0/3}
    {
        \pgfmathsetmacro{\colo}{\ColorTable[\c]};
        \pgfmathsetmacro{\opacity}{\TransparencyTable[\c]};
        \pgfmathsetmacro{\pat}{\PatternTable[\c]};
        \pgfmathsetmacro{\patcolo}{\PatternColorTable[\c]};
        \fill[\opacity,\colo] (3*\offsetscale*\width+\o*\width+\woffset,-.2*\y*\height) +(0.,0.) rectangle ++(\x*\width,-.2*\height);
        \fill[pattern=\pat, pattern color=\patcolo] (3*\offsetscale*\width+\o*\width+\woffset,-.2*\y*\height) +(0.,0.) rectangle ++(\x*\width,-.2*\height);
        \draw (3*\offsetscale*\width+\o*\width+\woffset,-.2*\y*\height) +(0.,0.) rectangle ++(\x*\width,-.2*\height);
    }
    \foreach \y/\x/\o/\c in {
                       1/1/0/4,
                       3/1/0/4,
                       4/.3/.7/4}
    {
        \pgfmathsetmacro{\colo}{\ColorTable[\c]};
        \pgfmathsetmacro{\opacity}{\TransparencyTable[\c]};
        \pgfmathsetmacro{\pat}{\PatternTable[\c]};
        \pgfmathsetmacro{\patcolo}{\PatternColorTable[\c]};
        \fill[\opacity,\colo] (3*\offsetscale*\width+\o*\width+\woffset,-.2*\y*\height) +(0.,0.) rectangle ++(\x*\width,-.2*\height);
        \fill[pattern=\pat, pattern color=\patcolo] (3*\offsetscale*\width+\o*\width+\woffset,-.2*\y*\height) +(0.,0.) rectangle ++(\x*\width,-.2*\height);
        \draw (3*\offsetscale*\width+\o*\width+\woffset,-.2*\y*\height) +(0.,0.) rectangle ++(\x*\width,-.2*\height);
    }

    \draw (.5*\width,.2*\height) node{$\gs{0}$};
    \draw (1.5*\offsetscale*\width-0.5*\woffset,.2*\height) node{$\gs{1}$};
    \draw (2.5*\offsetscale*\width-0.5*\woffset,.2*\height) node{$\gs{0,1}$};
    \draw (3.5*\offsetscale*\width+0.5*\woffset,.2*\height) node{$\pgs{0,1}$};
    \draw (-.25*\offsetscale*\width,-.1*\height) node{Ch 0:};
    \draw (-.25*\offsetscale*\width,-.3*\height) node{Ch 1:};
    \draw (-.25*\offsetscale*\width,-.5*\height) node{Ch 2:};
    \draw (-.25*\offsetscale*\width,-.7*\height) node{Ch 3:};
    \draw (-.25*\offsetscale*\width,-.9*\height) node{Ch 4:};
\end{tikzpicture}
    \begin{tikzpicture}[scale=0.9]
    \tikzstyle{every node}=[font=\scriptsize]
    \def\height{1.75}
    \def\width{1.75}
    \def\offsetscale{1.2}
    \def\shallowwidth{.7}
    \def\fullwidth{1}
    \def\stackhoffsetm{2.25}
    \def\coloone{"semitransparent,white!100!black"}
    \def\colotwo{"semitransparent,blue!60!black"}
    \def\colothree{"semitransparent,red!60!black"}
    \def\colofour{"semitransparent,green!60!black"}
    \def\colofive{"semitransparent,yellow!80!black"}
    \def\ColorTable{{"white!100!black",
                     "blue!60!black",
                     "yellow!100!black",
                     "cyan!100!white",
                     "magenta!100!black"}}
    \def\TransparencyTable{{"semitransparent",
                     "semitransparent",
                    "opaque",
                     "semitransparent",
                     "semitransparent",
                     "semitransparent",
                     "opaque"}}
    \def\PatternTable{{"north eastish lines",
                     "north westish lines",
                     "north westish lines",
                     "north eastish lines",
                     "dots",
                     "hatchish lines",
                     "north west lines, pattern color=black",
                     "north west lines, pattern color=black",
                     "north west lines, pattern color=black",
                     "north west lines, pattern color=black",
                     "semitransparent,yellow!80!black"}}
    \def\PatternColorTable{{"white",
                     "black",
                     "black",
                     "black",
                     "black",
                     "yellow!80!black"}}
    \def\GateInit{{0/1,1/.7,2/.1,3/.1,4/.1,5/.4,6/.1,7/.1}}
    \def\GatePad{{0/1/0,
                  1/.7/0,1/.3/.7,
                  2/1/0,
                  3/1/0,
                  4/1/0,
                  5/.4/0,5/.6/.4,
                  6/1/0,
                  7/1/0}}

    \pgfmathsetmacro{\woffset}{.5*(\fullwidth-\shallowwidth+\offsetscale-1)*\width};

    \foreach \y/\x/\o/\c in {0/1/0/2,
                       2/1/0/2,
                       4/.7/0/3}
    {
        \pgfmathsetmacro{\colo}{\ColorTable[\c]};
        \pgfmathsetmacro{\opacity}{\TransparencyTable[\c]};
        \pgfmathsetmacro{\pat}{\PatternTable[\c]};
        \pgfmathsetmacro{\patcolo}{\PatternColorTable[\c]};
        \fill[\opacity,\colo] (0*\offsetscale*\width+\o*\width+0*\woffset,-.2*\y*\height) +(0.,0.) rectangle ++(\x*\width,-.2*\height);
        \fill[pattern=\pat, pattern color=\patcolo] (0*\offsetscale*\width+\o*\width+0*\woffset,-.2*\y*\height) +(0.,0.) rectangle ++(\x*\width,-.2*\height);
        \draw (0*\offsetscale*\width+\o*\width+0*\woffset,-.2*\y*\height) +(0.,0.) rectangle ++(\x*\width,-.2*\height);
    }
    \foreach \y/\x/\o/\c in {
                       1/1/0/4,
                       3/1/0/4,
                       4/.3/.7/4}
    {
        \pgfmathsetmacro{\colo}{\ColorTable[\c]};
        \pgfmathsetmacro{\opacity}{\TransparencyTable[\c]};
        \pgfmathsetmacro{\pat}{\PatternTable[\c]};
        \pgfmathsetmacro{\patcolo}{\PatternColorTable[\c]};
        \fill[\opacity,\colo] (0*\offsetscale*\width+\o*\width+0*\woffset,-.2*\y*\height) +(0.,0.) rectangle ++(\x*\width,-.2*\height);
        \fill[pattern=\pat, pattern color=\patcolo] (0*\offsetscale*\width+\o*\width+0*\woffset,-.2*\y*\height) +(0.,0.) rectangle ++(\x*\width,-.2*\height);
        \draw (0*\offsetscale*\width+\o*\width+0*\woffset,-.2*\y*\height) +(0.,0.) rectangle ++(\x*\width,-.2*\height);
    }
    \draw (1*\offsetscale*\width-.15*\woffset,-.2*2.5*\height) node{$\concat$};
    \foreach \y/\x/\o/\c in {0/.7/0/3,
                       4/.7/0/3}
    {
        \pgfmathsetmacro{\colo}{\ColorTable[\c]};
        \pgfmathsetmacro{\opacity}{\TransparencyTable[\c]};
        \pgfmathsetmacro{\pat}{\PatternTable[\c]};
        \pgfmathsetmacro{\patcolo}{\PatternColorTable[\c]};
        \fill[\opacity,\colo] (1*\offsetscale*\width+\o*\width+0.5*\woffset,-.2*\y*\height) +(0.,0.) rectangle ++(\x*\width,-.2*\height);
        \fill[pattern=\pat, pattern color=\patcolo] (1*\offsetscale*\width+\o*\width+0.5*\woffset,-.2*\y*\height) +(0.,0.) rectangle ++(\x*\width,-.2*\height);
        \draw (1*\offsetscale*\width+\o*\width+0.5*\woffset,-.2*\y*\height) +(0.,0.) rectangle ++(\x*\width,-.2*\height);
    }
    \foreach \y/\x/\o/\c in {
                       1/.7/0/4,
                       2/.7/0/4,
                       3/.7/0/4}
    {
        \pgfmathsetmacro{\colo}{\ColorTable[\c]};
        \pgfmathsetmacro{\opacity}{\TransparencyTable[\c]};
        \pgfmathsetmacro{\pat}{\PatternTable[\c]};
        \pgfmathsetmacro{\patcolo}{\PatternColorTable[\c]};
        \fill[\opacity,\colo] (1*\offsetscale*\width+\o*\width+.5*\woffset,-.2*\y*\height) +(0.,0.) rectangle ++(\x*\width,-.2*\height);
        \fill[pattern=\pat, pattern color=\patcolo] (1*\offsetscale*\width+\o*\width+.5*\woffset,-.2*\y*\height) +(0.,0.) rectangle ++(\x*\width,-.2*\height);
        \draw (1*\offsetscale*\width+\o*\width+.5*\woffset,-.2*\y*\height) +(0.,0.) rectangle ++(\x*\width,-.2*\height);
    }
    \draw (2*\offsetscale*\width+.3*\woffset,-.2*2.5*\height) node{$\concat\;\ldots\;\Rightarrow$};

    \foreach \y/\x/\o/\c in {-.6/1/0/2,
                       2/1/0/2,
                       4.6/.7/0/3}
    {
        \pgfmathsetmacro{\colo}{\ColorTable[\c]};
        \pgfmathsetmacro{\opacity}{\TransparencyTable[\c]};
        \pgfmathsetmacro{\pat}{\PatternTable[\c]};
        \pgfmathsetmacro{\patcolo}{\PatternColorTable[\c]};
        \fill[\opacity,\colo] (2*\offsetscale*\width+\o*\width+\stackhoffsetm*\woffset,-.2*\y*\height) +(0.,0.) rectangle ++(\x*\width,-.2*\height);
        \fill[pattern=\pat, pattern color=\patcolo] (2*\offsetscale*\width+\o*\width+\stackhoffsetm*\woffset,-.2*\y*\height) +(0.,0.) rectangle ++(\x*\width,-.2*\height);
        \draw (2*\offsetscale*\width+\o*\width+\stackhoffsetm*\woffset,-.2*\y*\height) +(0.,0.) rectangle ++(\x*\width,-.2*\height);
    }
    \foreach \y/\x/\o/\c in {
                       .7/1/0/4,
                       3.3/1/0/4,
                       4.6/.3/.7/4}
    {
        \pgfmathsetmacro{\colo}{\ColorTable[\c]};
        \pgfmathsetmacro{\opacity}{\TransparencyTable[\c]};
        \pgfmathsetmacro{\pat}{\PatternTable[\c]};
        \pgfmathsetmacro{\patcolo}{\PatternColorTable[\c]};
        \fill[\opacity,\colo] (2*\offsetscale*\width+\o*\width+\stackhoffsetm*\woffset,-.2*\y*\height) +(0.,0.) rectangle ++(\x*\width,-.2*\height);
        \fill[pattern=\pat, pattern color=\patcolo] (2*\offsetscale*\width+\o*\width+\stackhoffsetm*\woffset,-.2*\y*\height) +(0.,0.) rectangle ++(\x*\width,-.2*\height);
        \draw (2*\offsetscale*\width+\o*\width+\stackhoffsetm*\woffset,-.2*\y*\height) +(0.,0.) rectangle ++(\x*\width,-.2*\height);
    }
    \foreach \y/\x/\o/\c in {-.6/.7/1/3,
                       4.6/.7/1/3}
    {
        \pgfmathsetmacro{\colo}{\ColorTable[\c]};
        \pgfmathsetmacro{\opacity}{\TransparencyTable[\c]};
        \pgfmathsetmacro{\pat}{\PatternTable[\c]};
        \pgfmathsetmacro{\patcolo}{\PatternColorTable[\c]};
        \fill[\opacity,\colo] (2*\offsetscale*\width+\o*\width+\stackhoffsetm*\woffset,-.2*\y*\height) +(0.,0.) rectangle ++(\x*\width,-.2*\height);
        \fill[pattern=\pat, pattern color=\patcolo] (2*\offsetscale*\width+\o*\width+\stackhoffsetm*\woffset,-.2*\y*\height) +(0.,0.) rectangle ++(\x*\width,-.2*\height);
        \draw (2*\offsetscale*\width+\o*\width+\stackhoffsetm*\woffset,-.2*\y*\height) +(0.,0.) rectangle ++(\x*\width,-.2*\height);
    }
    \foreach \y/\x/\o/\c in {
                       .7/.7/1/4,
                       2/.7/1/4,
                       3.3/.7/1/4}
    {
        \pgfmathsetmacro{\colo}{\ColorTable[\c]};
        \pgfmathsetmacro{\opacity}{\TransparencyTable[\c]};
        \pgfmathsetmacro{\pat}{\PatternTable[\c]};
        \pgfmathsetmacro{\patcolo}{\PatternColorTable[\c]};
        \fill[\opacity,\colo] (2*\offsetscale*\width+\o*\width+\stackhoffsetm*\woffset,-.2*\y*\height) +(0.,0.) rectangle ++(\x*\width,-.2*\height);
        \fill[pattern=\pat, pattern color=\patcolo] (2*\offsetscale*\width+\o*\width+\stackhoffsetm*\woffset,-.2*\y*\height) +(0.,0.) rectangle ++(\x*\width,-.2*\height);
        \draw (2*\offsetscale*\width+\o*\width+\stackhoffsetm*\woffset,-.2*\y*\height) +(0.,0.) rectangle ++(\x*\width,-.2*\height);
    }

    \draw (.5*\width,.2*\height) node{$\pgs{0,1}$};
    \draw (1.5*\offsetscale*\width-0.5*\woffset,.2*\height) node{$\pgs{1}$};
    \draw (3.0*\offsetscale*\width+0.7*\woffset,.26*\height) node{Concatenated Data};
    \draw (4.15*\offsetscale*\width,.025*\height) node{$\rightarrow\text{GS}_0$};
    \draw (4.15*\offsetscale*\width,-.27*\height+.025*\height) node{$\rightarrow\text{GS}_1$};
    \draw (4.15*\offsetscale*\width,-.5*\height+.0*\height) node{$\rightarrow\text{GS}_2$};
    \draw (4.15*\offsetscale*\width,-.76*\height+.0*\height) node{$\rightarrow\text{GS}_3$};
    \draw (4.15*\offsetscale*\width,-1.02*\height+.0*\height) node{$\rightarrow\text{GS}_4$};
\end{tikzpicture}
    \caption{Merging (top) and concatenation (bottom) of gate slices. Each gate is broken into its constituent pulses, separated by output channel, and assigned a numeric index, $n$, to form its corresponding gate slice, $\gs{n}$. When gates are run in parallel, their gate slice data are merged on a per-channel basis. If two gate slices define pulses on the same channel, they are merged as long as the pulses are either identical or, for pulses of different duration, identical over the duration of the shorter pulse (in which case the longer pulse is used). Gate slices are then padded, $\gs{n}\pad\pgs{n}$, with NOP pulses to ensure that data is appropriately scheduled for alignment of pulse data for subsequent gates. After gate slices are merged and padded, they are concatenated for back-to-back execution and the processed data for each channel is uploaded to its corresponding gate sequencer, $\text{GS}_\text{ch}$.
    }
    \label{gateSliceMerge}
\end{figure}

Gate slices come in two basic forms: unpadded and padded, represented by $\mathfrak{S}$ and $\mathfrak{P}$ respectively. Unpadded gate slices support merging, $\gs{0}\merge\gs{1}$, and padding, $\gs{n}\pad\pgs{m}$, while padded gate slices support concatenation, $\pgs{0}\concat\pgs{1}$. 

Merging gate slices requires extra consideration. Gates comprising mutually independent channel sets are trivially merged, however gates which employ a common channel might be compatible. This might happen for parallel counter-propagating gates, which always use the global beam. If the data is identical on the global beam, then the gates can be merged, but additional checks are required for conflicting data. Because gates might have mismatched durations, but otherwise equivalent data on a shared channel, merge operations are considered valid if the gate data are identical up to the shorter duration, in which case the longer segment is used, as shown in Fig.~\ref{gateSliceMerge}.

Once all parallel gate slices are merged, each gate slice is padded to form the final representation for gate data to be stored in the LUTs, at which point they are appropriately packed and encoded for programming. Since the TIR already compresses unique gate calls, initial programming data is determined from a single pass over the TIR's gate and parallel block tables.  Padded gate slices ultimately comprise a collection of pulse handles, generated by the pulse managers, and are assigned indices matching their respective parallel and gate blocks in the TIR. 

References to the lowest-level blocks in the TIR now point to padded gate slices, and a full walk of the TIR is used for fully concatenating them.  Redundant calls (e.g.  $\pgs{0}\concat\pgs{1}\concat\pgs{0}\concat\ldots$) finally come into play and loops are expanded via repeated concatenation.  Because of the tight correspondence of the padded gate slices with the pulse managers, and the pulse managers with the LUTs, concatenation essentially constructs the full set of GLUT addresses for a gate sequence, which are packed and encoded in preparation for transfer to PL.

This final representation can be easily stored in RAM for subsequent access and exposed to PL or the RPU, where over $10^7$ gates can fit into 128 MB. While circuit depths typically don't come close to this limit, certain characterization protocols, such as gate set tomography (GST)~\cite{nielsen2020gate}, can contain thousands of subcircuits. Ideally, these circuits are sampled in random order, which is enabled by the ability to quickly access such a large number of sequences. 

\subsection{Branch handling}

Because of the simple mechanism for handling branching at the hardware level, efficient usage of the branching infrastructure is primarily a software problem.  Branch sequences must be able to handle every possible measurement outcome with a common set of sequence bytecode. The most trivial approach is to use GLUT addresses that start at zero, increment for each gate in a branch, and continue incrementing on subsequent branches. This limits the number of gates that can be run in branch sequences before partial reprogramming is required.

More efficient representations can be calculated by looking for correlations in gate redundancy across different cases within a branch, as well as across branches. However, the computing overhead tends to outweigh the benefit, since the branching infrastructure is designed for very fast ($\approx$ 50 ns) response for techniques such as error correction. Most of these use cases typically require a handful of gates per branch, so we have opted to implement the trivial approach until more exhaustive methods become necessary.

\subsection{Partial reprogramming and gate mutation}

Circuits with a lot of unique gate data may exceed the gate sequencer LUT capacities. Even in extreme cases where all channels are simultaneously running waveforms in which data is never reused, taking advantage of the LUTs is often beneficial. This is due to the disproportionately larger memory size of the URAM used for the LUTs compared to the total available storage of standard block RAM (BRAM) primitives typically used for FIFOs. If the rate at which pulselet words are consumed varies over the course of a circuit, staggered bursts of programming data can be strategically interleaved with sequence data to prevent FIFO underflows. 

Another case where partial reprogramming is of great utility is \emph{in-situ} gate mutations during high-level feedback on gate definitions. A standard use case is for shimming out slow drift in control parameters~\cite{Drift2020}. This involves interleaving standard circuits with specially-designed calibration circuits used to probe a particular error. Gate definitions are then recalculated on a shot-to-shot basis according to individual measurement outcomes. 

The metric used for gate recompilation is set by the initial Doppler cooling stage during state preparation. Doppler cooling time, or 1 ms, is a good approximation for downtime between a measurement and the start of the next gate sequence\footnote{Other state preparation steps such as sideband cooling require the qubit laser, and are treated as an extension of a quantum circuit.}.  Adaptive cooling techniques are typically used, where the Doppler cooling stage is repeated until ion fluorescence exceeds a minimum threshold.  This is done to ensure that the ion temperatures are sufficiently low enough for sideband cooling, and also used to check for ion loss. The Doppler cooling stage is thus considered a suitable place for any processes that are not deterministically timed, since adaptive cooling already breaks timing determinism and excess cooling time doesn't have any negative impact.

Gate recalculations are triggered once a measurement is received, where specified gates are refetched and compiled. However, the compilation step for gate mutations makes use of the previously-cached definition to determine differences in the data and their corresponding PLUT addresses, to minimize recalculation and data transfer size. 

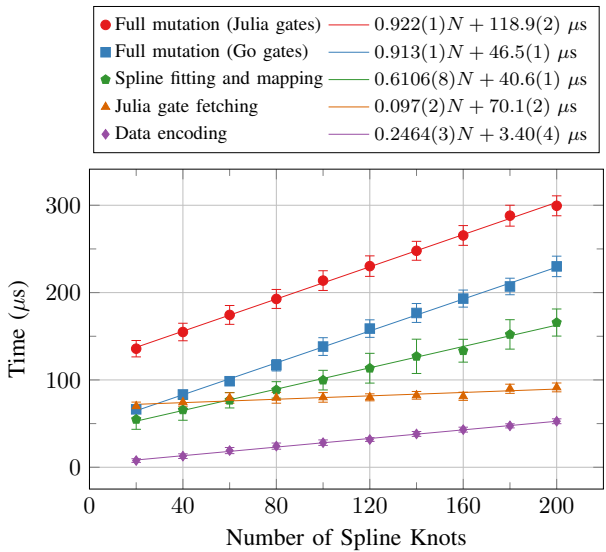
\begin{figure}
    \centering
    \scalebox{.9}{
    \begin{minipage}[t]{.85\textwidth}
        \vspace{0pt}
        \definecolor{d1}{rgb}{0.8941176470588236,0.10196078431372549,0.10980392156862745}
\definecolor{d2}{rgb}{0.21568627450980393,0.49411764705882355,0.7215686274509804}
\definecolor{d3}{rgb}{0.20196078431372547,0.5862745098039216,0.1901960784313726}
\definecolor{d4}{rgb}{0.85,0.3980392156862745,0.0}
\definecolor{d5}{rgb}{0.596078431372549,0.3058823529411765,0.6392156862745098}
\begin{tikzpicture}
    \pgfplotsset{    
    every axis x label/.style={
        at={(ticklabel* cs:1.05)},
        anchor=west,
    },
    every axis y label/.style={at={(current axis.above origin)},
    anchor=north east,
    yshift=1cm,},
    compat=1.17,
    every axis legend/.append style={
        legend columns=5,font=\footnotesize},
        width=\axisdefaultwidth*1.1,
    }
    \begin{axis}[
      cycle multiindex* list={
        d1,d2,d3,d4,d5\nextlist
        solid,{solid, mark options={solid}}\nextlist
        mark=*,mark=square*,mark=pentagon*,mark=triangle*,mark=diamond*
      },
      xtick={0,40,80,...,200},
      minor tick num=1,
      width=\axisdefaultwidth*0.9,
      height=\axisdefaultheight*.65,
      xmin=0,
      xlabel={Number of Spline Knots},
      ylabel={Time ($\mu$s)},
      grid=major,
      legend cell align=left,
      transpose legend,
      legend style={at={(0.5,1.05)},anchor=south},
      scale only axis,
    ]
      \addplot+ [only marks, error bars/.cd, y dir = both, y explicit] 
                table [col sep=comma,header=false,x index=0,y index=1,y error index=2] 
                    {figures/gnufctallcsv.csv};
      \addplot+ [only marks, error bars/.cd, y dir = both, y explicit] 
                table [col sep=comma,header=false,x index=0,y index=1,y error index=2] 
                    {figures/gnufcpgtallcsv.csv};
      \addplot+ [only marks, error bars/.cd, y dir = both, y explicit] 
                table [col sep=comma,header=false,x index=0,y index=1,y error index=2] 
                    {figures/gnucmptallfullcsv.csv};
      \addplot+ [only marks, error bars/.cd, y dir = both, y explicit] 
                table [col sep=comma,header=false,x index=0,y index=1,y error index=2] 
                    {figures/gnuftallfullcsv.csv};
      \addplot+ [only marks, error bars/.cd, y dir = both, y explicit] 
                table [col sep=comma,header=false,x index=0,y index=1,y error index=2] 
                    {figures/gnuetallfullcsv.csv};
      \addplot+ [mark=none,domain=20:200]{0.922*x + 118.9};
      \addplot+ [mark=none,domain=20:200]{0.913*x + 46.5};
      \addplot+ [mark=none,domain=20:200]{0.6106*x + 40.6};
      \addplot+ [mark=none,domain=20:200]{0.0972*x + 70.1};
      \addplot+ [mark=none,domain=20:200]{0.2464*x + 3.40};
      \legend{Full mutation (Julia gates),
              Full mutation (Go gates),
              Spline fitting and mapping,
              Julia gate fetching,
              Data encoding,
              $0.922(1) N + 118.9(2)~\mu$s,
              $0.913(1) N + 46.5(1)~\mu$s,
              $0.6106(8) N + 40.6(1)~\mu$s,
              $0.097(2) N + 70.1(2)~\mu$s,
              $0.2464(3)N + 3.40(4)~\mu$s}
    \end{axis}
\end{tikzpicture}
    \end{minipage}
    }
    \caption{On-chip gate mutation times as a function of the number of spline
    knots. Fetch time is the elapsed duration between issuing a gate request to
    Julia and receiving the gate definition, including conversion to a data
    structure used natively by the Go compiler code. Spline fitting and
    mapping includes calculation of natural cubic spline coefficients,
    subsequent mapping to the format used by the spline engines, and
    reregistration with the pulse managers, all of which is tightly coupled to the
    coefficient calculation. 
    Data encoding involves recasting the spline
    coefficients into a byte array with the appropriate LUT programming
    metadata. Full mutations are compared for gates defined in Julia and in Go,
    which is primarily offset by the on-chip fetch time. Each point is averaged
    over 10,000 repetitions, and plotted error bars are standard deviations.  
    }
    \label{mutationBenchmark}
\end{figure}

Performance of gate mutations, shown in Fig.~\ref{mutationBenchmark}, is more critical than other stages where reasonable downtime is typically expected.  However, the fact that common data is shared among gates can be used to one's advantage. Gates of similar type, such as amplitude-modulated single-qubit gates which only differ by a phase, might require updates to common parameters, such as frequency or amplitude. In this case, only one gate may require an update to affect the whole class. On the other hand, gate mutations can have unintended side effects if gate data is inadvertently changed. To avoid such side effects, pulses are tagged with a ``mutation id'' that can be used to artificially change the uniqueness during compilation. Thus, mutations can be applied to pulses with a particular mutation id, and this id can be shared for a class of gates for which mutations are intended.

\section{Discussion}

We have presented a coherent control system, capable of running nearly the entire software stack on chip. This system can compile quantum assembly, written in Jaqal, down to pulse-level gate descriptions which are subsequently converted to rf pulses delivered by the same device. Moreover, the system integrates a high-level interface for calculating complex waveform modulation parameters needed for advanced gate designs.

While performance requirements vary, the software was designed to be modular, allowing separate stages such as parsing, compilation, and gate definitions, to be optionally executed on or off chip. All elements of this modular architecture can be run simultaneously on different systems, allowing one to optionally parse \emph{off} chip and compile \emph{on} chip, or even parse \emph{on} chip and compile \emph{off} chip.
This modularity allows one to simply send the input to a particular machine (or target port) on a case-by-case basis.
Integrating configurable modularity seamlessly into the software design can be used to quickly switch between various modes of operation, such as calibration scans, algorithmic feedback for stabilizing against drift, or very-high-level algorithms which require more computational power than can be afforded on chip.

Our primary focus for performance benchmarks targeted drift control applications, where a combination of high-level classical computation and sub-millisecond turnaround time is necessary. This was tested on moderately complex gates with multiple parameters modulated via cubic splines. The gates were defined on chip using Julia to support flexible, dynamically-loadable code which is just-in-time compiled.

For gates that employ 20--150 spline knots, which is a typical range used by QSCOUT, the round trip times for a full gate mutation ranged from $\approx$140~$\mu$s--260~$\mu$s. For gates that are written in Go, and added directly to the compiler, this time is reduced overall by $\approx$~70~$\mu$s.  These results are comparable to typical two-qubit gate times of $\approx$~200~$\mu$s, and shorter than the Doppler-cooling stage of 1~ms.

By virtue of shared gate data in the compressed representation used in hardware, the ability to simultaneously modify multiple gates with a single definition can be used to effectively increase performance.  Given that compilation on the embedded system can outpace an external server when accounting for network transfer times, these results are an encouraging step towards a fully distributed system in which classical computing resources scale with channel number.

\section*{Acknowledgment}

We want to thank Peter Maunz, whose contributions to the initial Octet design were invaluable, as well as Andrew Landahl, Kenneth Rudinger, Antonio Russo, and Benjamin Morrison for their continuing development efforts of the Jaqal language.

This material is based upon work supported by the U.S. Department of Energy, Office of Science, Office of Advanced Scientific Computing Research under the Quantum Testbed Program and National Quantum Information Science Research Centers, Quantum Systems Accelerator.  Sandia National Laboratories is a multimission laboratory managed and operated by National Technology and Engineering Solutions of Sandia, LLC, a wholly owned subsidiary of Honeywell International Inc., for the U.S. Department of Energy's National Nuclear Security Administration under Contract DE-NA0003525. This article describes objective technical results and analysis.  Any subjective views or opinions that might be expressed in the article do not necessarily represent the views of the U.S. Department of Energy or the U. S.  Government.  SAND2022-10083 C

\bibliographystyle{IEEEtran}
\bibliography{main.bbl}

\end{document}